# Assets Forecasting with Feature Engineering and Transformation Methods for LightGBM

Konstantinos-Leonidas Bisdoulis
Model Lyceum of Anavryta, Marousi, Greece
e-mail: konbisdoulis@gmail.com

## Abstract

Fluctuations in the stock market rapidly shape the economic world and consumer markets, impacting millions of individuals. Hence, accurately forecasting it is essential for mitigating risks, including those associated with inactivity. Although research shows that hybrid models of Deep Learning (DL) and Machine Learning (ML) yield promising results, their computational requirements often exceed the capabilities of average personal computers, rendering them inaccessible to many. In order to address this challenge in this paper we optimize LightGBM (an efficient implementation of gradient-boosted decision trees (GBDT)) for maximum performance, while maintaining low computational requirements. We introduce novel feature engineering techniques including indicator-price slope ratios and differences of close and open prices divided by the corresponding 14-period Exponential Moving Average (EMA), designed to capture market dynamics and enhance predictive accuracy. Additionally, we test seven different feature and target variable transformation methods, including returns, logarithmic returns, EMA ratios and their standardized counterparts as well as EMA difference ratios, so as to identify the most effective ones weighing in both efficiency and accuracy. The results demonstrate Log Returns, Returns and EMA Difference Ratio constitute the best target variable transformation methods, with EMA ratios having a lower percentage of correct directional forecasts, and standardized versions of target variable transformations requiring significantly more training time. Moreover, the introduced features demonstrate high feature importance in predictive performance across all target variable transformation methods. This study highlights an accessible, computationally efficient approach to stock market forecasting using LightGBM, making advanced forecasting techniques more widely attainable.

## 1. Introduction

The stock market exerts a profound influence on the financial well-being of individuals and societies alike. Even though its effects are directly manifest in corporations and investors, its reach stretches far beyond, entrenching into everyday life in the form of consumer prices, economic policies and inflation. While its growth is indicative of economic prosperity, its demise is often translated into repercussions for both participants and non-participants. It is therefore imperative that individuals pre-emptively act to mitigate such consequences.

Time series forecasting constitutes an established approach for tackling the nuances of the stock market. It refers to the use of scientific methods in order to predict future values of a time-dependent variable based on historical data. Over the past decades, it has been extensively applied to the stock market as it allows for informed decision-making, economic planning and, potentially, passive income.

As technological growth continued its exponential surge, innovation in the field of time series forecasting accelerated. Initially statistical methods dominated the field, but as artificial intelligence rapidly advanced ML and DL were applied to financial time series such as the stock market. Their ability to effectively deal with complex and non-linear data proved revolutionary, with numerous contemporary studies postulating hybrid models of ML and DL. While highly accurate, the computational requirements of those models often restrict their utility to institutions with significant technological resources, thereby limiting their accessibility to individual investors, independent researchers, small-scale practitioners and ordinary people.

To address this challenge this study focuses on LightGBM, a ML model that uses tree-based learning algorithms [1]. Unlike heavier Deep Learning models, LightGBM is computationally efficient and highly scalable which renders it feasible for broader adoption. It has shown impressive results in time series forecasting as made evident by the M5 Forecasting Competition [2]. In this study we aim to further enhance its performance by introducing novel feature engineering techniques and identifying the optimal feature and target variable transformation methods.

## 2. Related Work

### *2.1 Statistical Models*

Statistical models have historically been widely used in the field of financial time series forecasting. Among these, the Auto Regressive Moving Average (ARIMA) model produces forecasts based on identification of linear correlations in the data. Consequently, it has demonstrated high accuracy in areas like energy forecasting [3], and hotel price forecasting [4]. However, due to its limitations of solely modeling linear relationships it falls short when it comes to non-linear heteroskedastic time series like the stock market [5]. As a result, Generalized Autoregressive Conditional Heteroskedasticity (GARCH) models were developed, successfully addressing this issue by directly modeling volatility. As Farah Hayati Mustapa et al. showed, a hybrid GARCH-ARIMA model managed to achieve limited accuracy (R-squared=0.023910) in forecasting the S&P 500 index [7].

### *2.2 Machine Learning Models*

Emerging at the start of the century, ML models capitalized on technological improvements to be effectively applied to the stock market. Excelling at deciphering the intricate relationships of multi-variate datasets with features that provide additional information, ML models proved more reliable at dealing with the non-linear nature of the stock market. Notable ML models include Support Vector Machines, Random Forests and Gradient Boosting Machines with the latter being implemented more efficiently in models like

LightGBM, XGBoost, CatBoost and AdaBoost. Ma shangchen et al. showed that the LightGBM and GBDT models both managed to produce higher returns than the S&P 500 index in the same period, with LightGBM achieving up to 394% total returns when used to forecast the index for the period of 2016 to 2018 [6].

### *2.3 Deep Learning Models*

Building upon the success of ML models, DL models specialize in modeling temporal dependencies in a time series and possess the ability of extracting informative features directly from the target variable. However, their adoption is more sluggish, in part due to their immense requirements for adequate training. Some of those include sufficiently large datasets; ample Random Access Memory (RAM); and highly parallelizable processors like state-of-the-art GPUs and TPUs. Hybrid ML-DL models have been shown to be particularly effective at time series forecasting, leveraging the strengths of both technologies. Yuankai Guo et al. postulated an LSTM-LightGBM hybrid model, predicting the values of features like volume and closing price with LSTM which were then used by LightGBM to make the final prediction. Their proposed model outperformed standalone Long Short Term Memory (LSTM) and Recurrent Neural Network (RNN) models [8].

## 3. Our Contribution

In this study we seek to optimize the performance of LightGBM, shifting away from the reliance on powerful hardware typically required for Deep Learning models. Seven different target variable transformations are systematically evaluated with training being conducted on a feature set containing different variants of features, each created with one of those same transformation methods. Moreover, we introduce novel cross features such as slope differences between price and indicators and open-previous close price differences so as to enhance accuracy without significantly increasing training time. Through this approach we seek to empower a broader audience to harness the benefits of machine learning in finance, contributing to a more inclusive and democratized financial forecasting landscape, which can in turn lead to more innovation.

## 4. LightGBM

The Light Gradient Boosting Machine, or LightGBM for short, is an advanced, open-source machine learning framework which builds on GBDTs, markedly improving efficiency. GBDTs make forecasts by iteratively training decision trees on the data to correct errors from previous trees. More specifically, a base prediction, which can be the mean of the target variable, is initially used to forecast the data. Subsequently, decision trees are fitted to the residuals of the base prediction in order to predict the errors. This process, called a boosting round, repeats until accuracy evaluated on a validation set doesn’t improve for a specified number of boosting rounds.

LightGBM improves GBDTs in several distinct ways, among others including:

i. **Gradient One-Side Sampling (GOSS):**
Gradients are the derivative of the loss function. They indicate how much the residuals change in relation to the model parameters. GOSS places higher focus on instances with larger gradients as they are more informative to the model, while keeping a random subset of instances with lower gradients as well. Through this process, higher efficiency is achieved without significant compromise in accuracy.

ii. **Histogram-Based Binning:**
Binning combines multiple instances of values of features into one discrete bin. A bin is of the range of values and number of data points it consists of. Gradients are calculated on bins rather than individual data points, resulting in lower memory usage and faster training. Additionally, this approach filters noisy data, thereby promoting generalization.

iii. **Leaf-Wise Growth:**
LightGBM grows decision trees leaf-wise unlike GBDTs which use level-wise growth. In level-wise growth all nodes at the last level of a decision tree are split simultaneously, with the number of nodes increasing exponentially. In contrast, leaf wise growth only splits the most informative nodes regardless of level, creating deeper more complex trees and lessening computational strain.

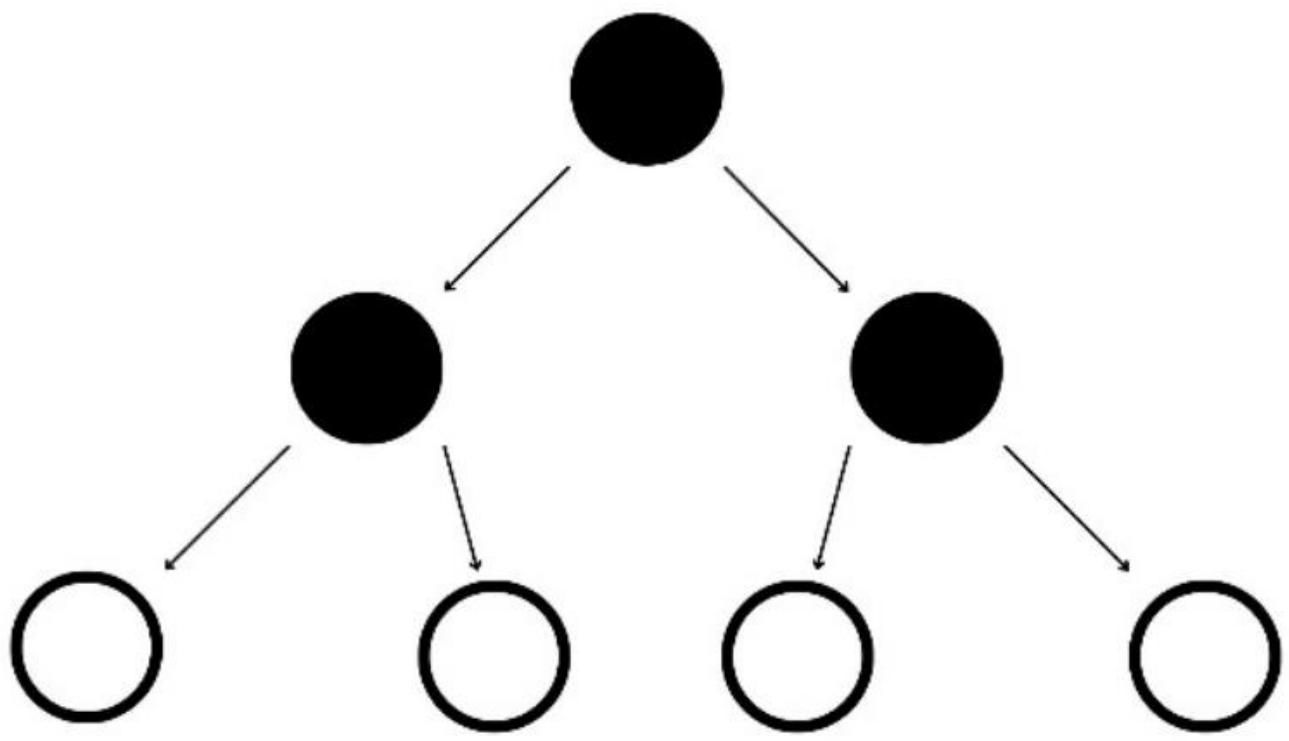

*Figure 1: Level-Wise Growth*

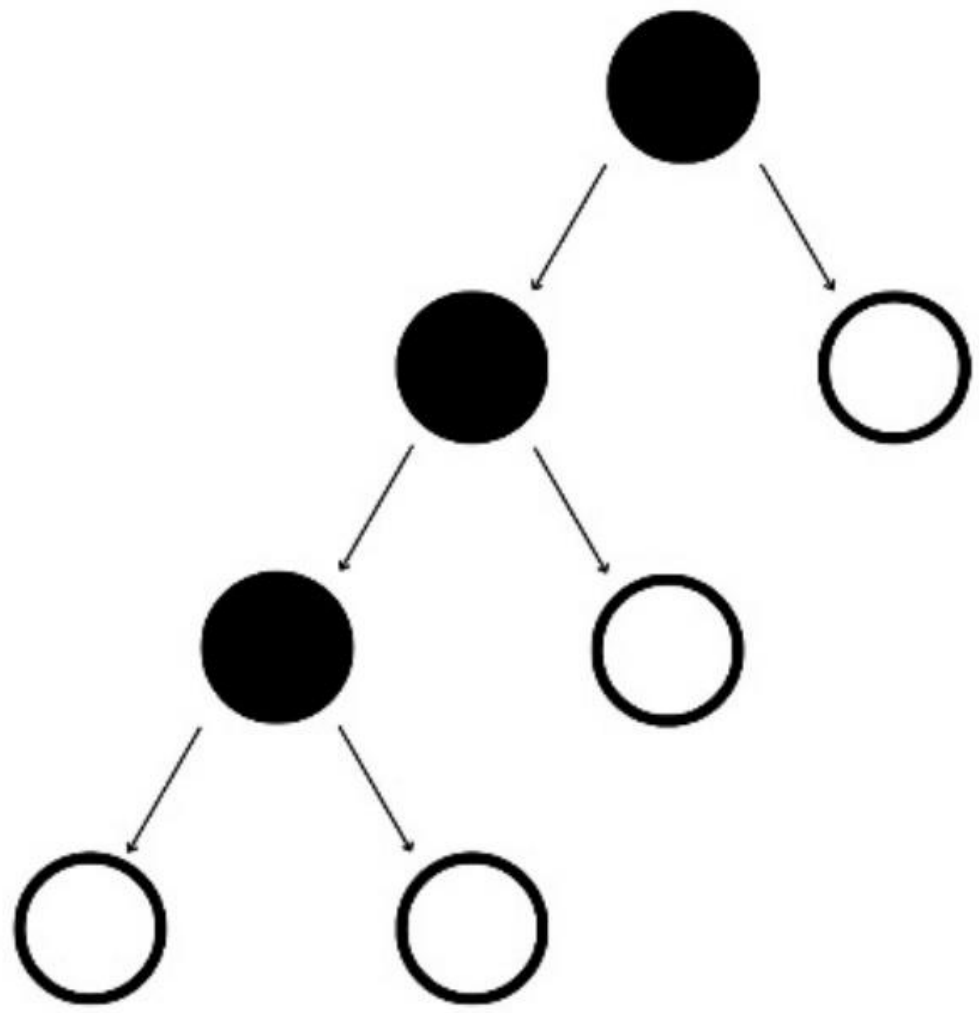

*Figure 2: Leaf-Wise Growth*

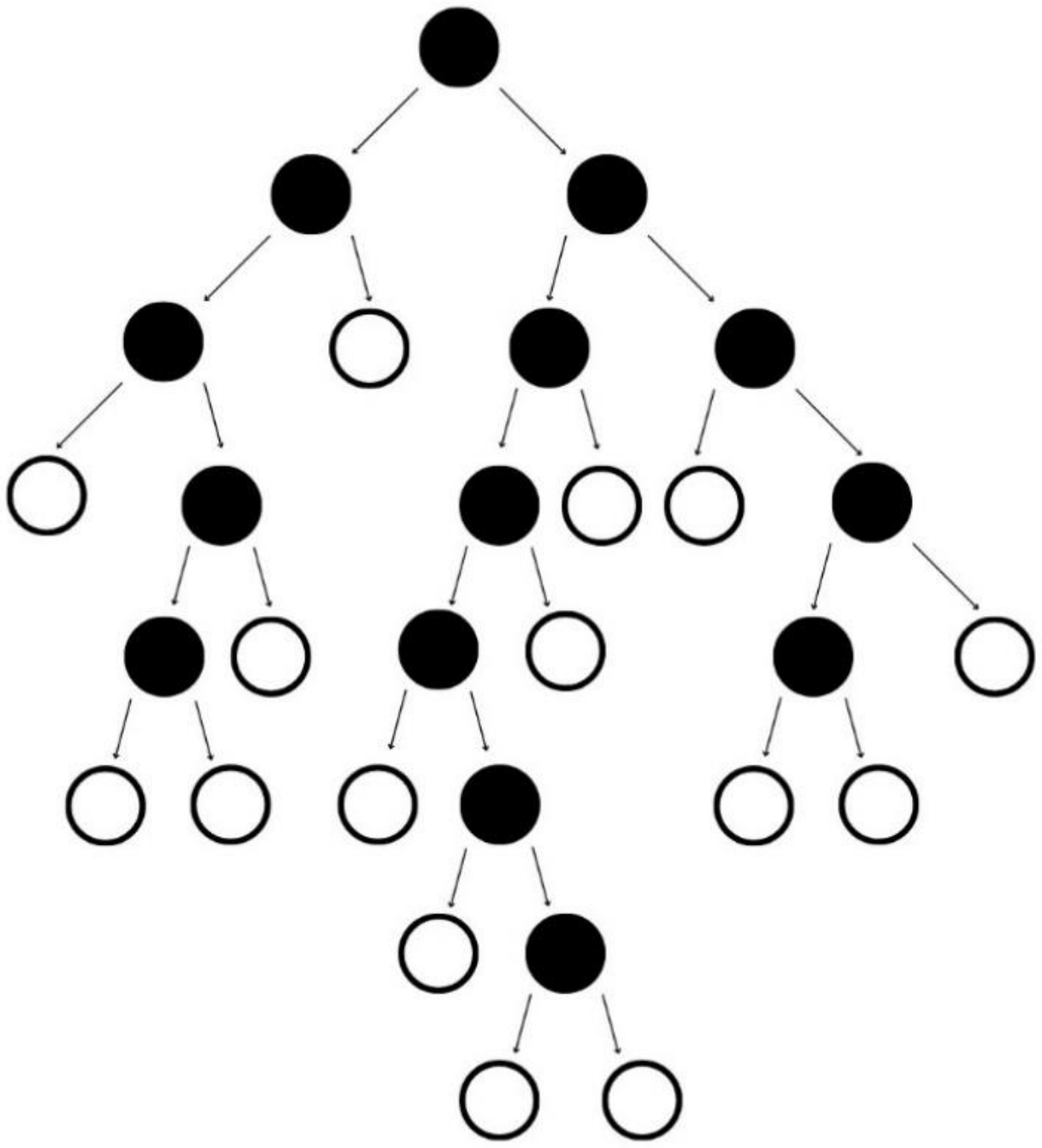

*Figure 3: Example of a complex tree grown leaf-wise*

## 5. Methodology

### *5.1 Data Collection*

The fluctuations of the Apple stock price (AAPL) are generally regarded analogous to that of blue-chip stocks, as in they're amply liquid, relatively involatile and stable. As a constituent of major indices like the S&P 500, AAPL also reflects broader market trends, making it a relevant case for evaluating stock market forecasting methodologies. Therefore, we opt for AAPL to conduct testing. The open, high, low and close (OHLC) prices and volume are extracted from TradingView's export chart data feature, with a total of 8137 data points and dates spanning from May 1992 to September 2024 on the daily timeframe. TradingView uses data from Cboe One, which accounts for approximately 10% of the US equities market share [9]. The days on which the market was closed are not part of the dataset and thus no missing values are present.

### *5.2 Feature Engineering*

We will be forecasting the close price (target variable) of a given day at the opening time of the market, allowing for modeling of overnight price gaps and the use of the same day's open price as a feature. We calculate a plethora of both typical and novel features from the OHLC prices and volume, aiming to model different intricate market behaviors and dynamics, often borrowing concepts from Technical Analysis (TA) and Statistics.

*Note: For clarity, we will refer to the open price of the same day as the target close price simply as open. For the open, high, low and close which correspond to the previous day of trading activity relative to the target variable we will be referring to them as openprev, highprev, lowprev and closeprev respectively.*

### i. Typical price

Typical price is calculated using the formula:

$$\text{Typical Price} = \frac{\text{High} + \text{Low} + \text{Close}}{3}$$

*Equation 1: Typical price*

It provides the overall sentiment for each trading day.

### ii. Lag features

When generating predictions, LightGBM only considers values of features at the corresponding row of data. Time series forecasting by definition uses historical data to predict a future value. It is essential, therefore, that we use past values of price as features to adequately inform the model. We use lag periods of 1, 5 and 30 for open, closeprev and typical to get information about the previous day, the same day of the previous week and the day placed at 30 weekdays before, or approximately one and half month.

### iii. Statistical properties

A rolling mean, or otherwise known as Simple Moving Average (SMA), can provide additional insight into market trends. Rolling standard deviation and min/max are also utilized in certain features including volume, open, closeprev and typical price.

### iv. Technical Indicators

Research has shown that technical indicators can be particularly informative in quantitative finance [10]. Specifically, indicators like the Psychological Line indicator (PSY), Stochastic %K and %D, Moving Average Convergence Divergence indicator (MACD), Commodity Channel Index (CCI) and Chande Momentum Oscillator (CMO) have ranked high in feature importance in previous studies [7]. In addition to the aforementioned, we make use of the Relative Strength Index (RSI), Chaikin Volatility, Rate of Change (ROC), Average True Range (ATR) and Exponential Moving Average (EMA). All indicators are calculated based on closeprev. The following table shows the modeling property of each indicator.

| ***Property*** | ***Indicators*** |
|---|---|
| *Momentum* | ROC, RSI, CMO, MACD, Stochastic %K and %D |
| *Volatility* | ATR, Chaikin Volatility |
| *Trend* | EMA, SMA |

*Table 1: Modeling Property of each indicator*

v. **Timestamp Processing**

We use cyclical features to model the periodicity of trading days. Day of the week, day of the month and month of the year are represented by sine functions, with the goal of training the model to identify seasonal patterns.

vi. **Cross Features**

Cross Features refer to the combination of existing for the creation of new ones, so as to capture relationships between variables. We calculate the difference of open with closeprev which represents the overnight price gap. Moreover, we compute the difference of open with closeprev lag 1 which indicates the total distance traveled by price over one day and two nights. Lastly, the ratio of ATR relative to open is utilized, essentially stationarizing ATR in an unconventional way, since generally as price increases so does volatility.

vii. **Slope Differences - Fixed and Dynamic**

The case of an indicator moving in the opposite direction with price is classified as a divergence in TA. Divergencies can provide an alternative perspective on the future direction of price which can be very informative. To model divergencies in the context of a regression LightGBM, we opt to take the difference of the slope of the specified indicator with the slope of price over a fixed period, effectively capturing how price changes in relation to an indicator. On top of the fixed period method, we create dynamically adjusted periods with the help of the ZigZag indicator. The ZigZag indicator identifies past trend reversal points which are used to identify the length of current trend and use it as a period for the slope difference. In other words, we study the divergencies within the boundaries of the trend price finds itself at a given time. We use those divergencies with the ROC of closeprev relative to closeprev, the volume relative to closeprev and PSY relative to closeprev.

### *5.3 Data Processing*

As LightGBM uses a tree-based algorithm, the input data needs to be stationary. That is because tree-based models rely on splitting data into subsets based on feature thresholds. In order to make predictions the model identifies in which of those subsets the current features belong in and based on the values of the past target variables in those subsets it generates a prediction. If features or the target variable exhibit changes in variance (heteroskedasticity) or mean (trend), then the model will struggle to identify in which subset they belong as the current values will have significantly shifted from past values. Stationarity resolves this issue by ensuring that statistical properties like mean and variance remain constant over time, allowing the model to effectively generalize.

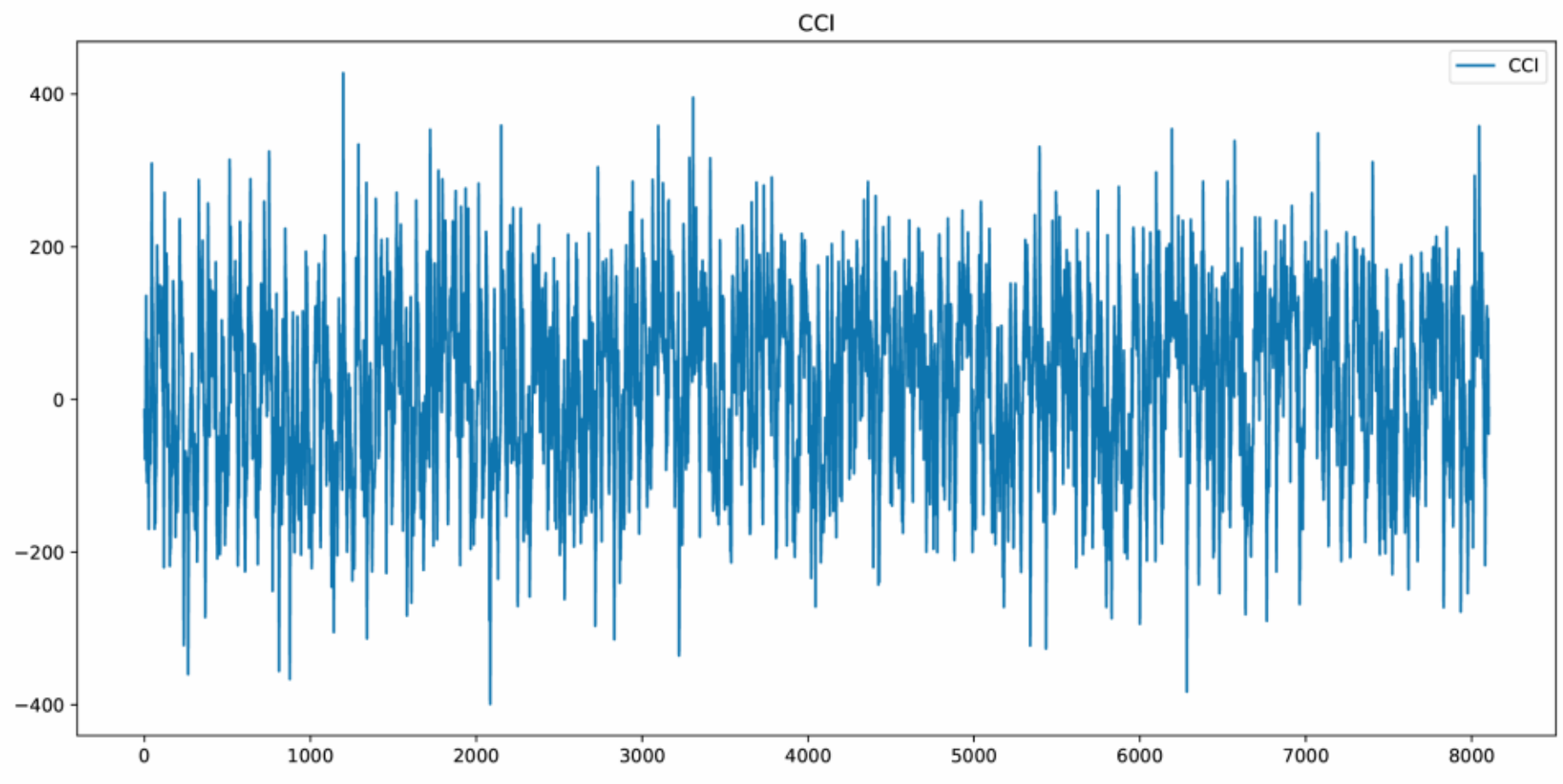

*Figure 4: Example of stationary time series, the CCI*

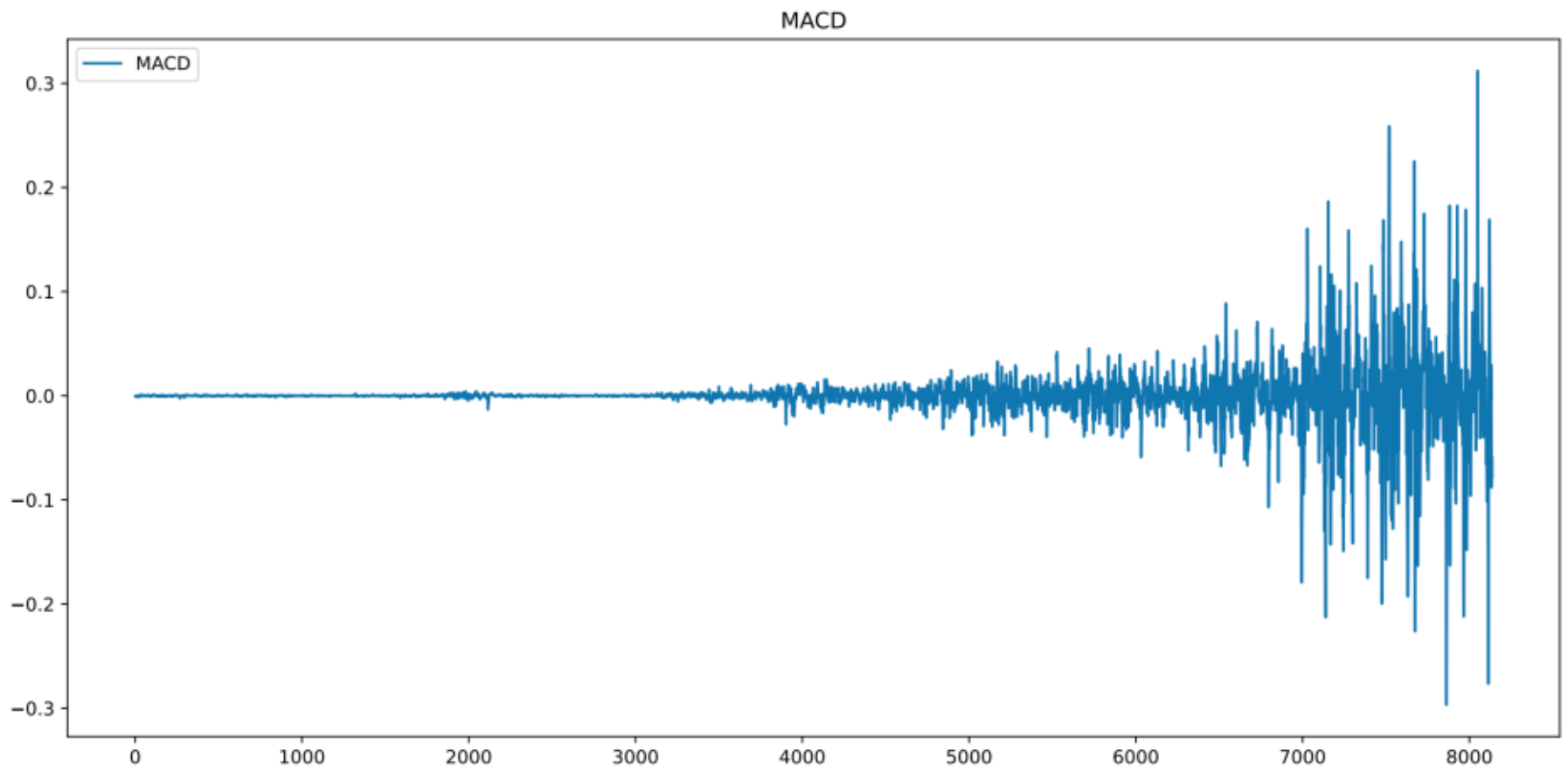

*Figure 5: Example of heteroskedastic, non-stationary time series, the MACD*

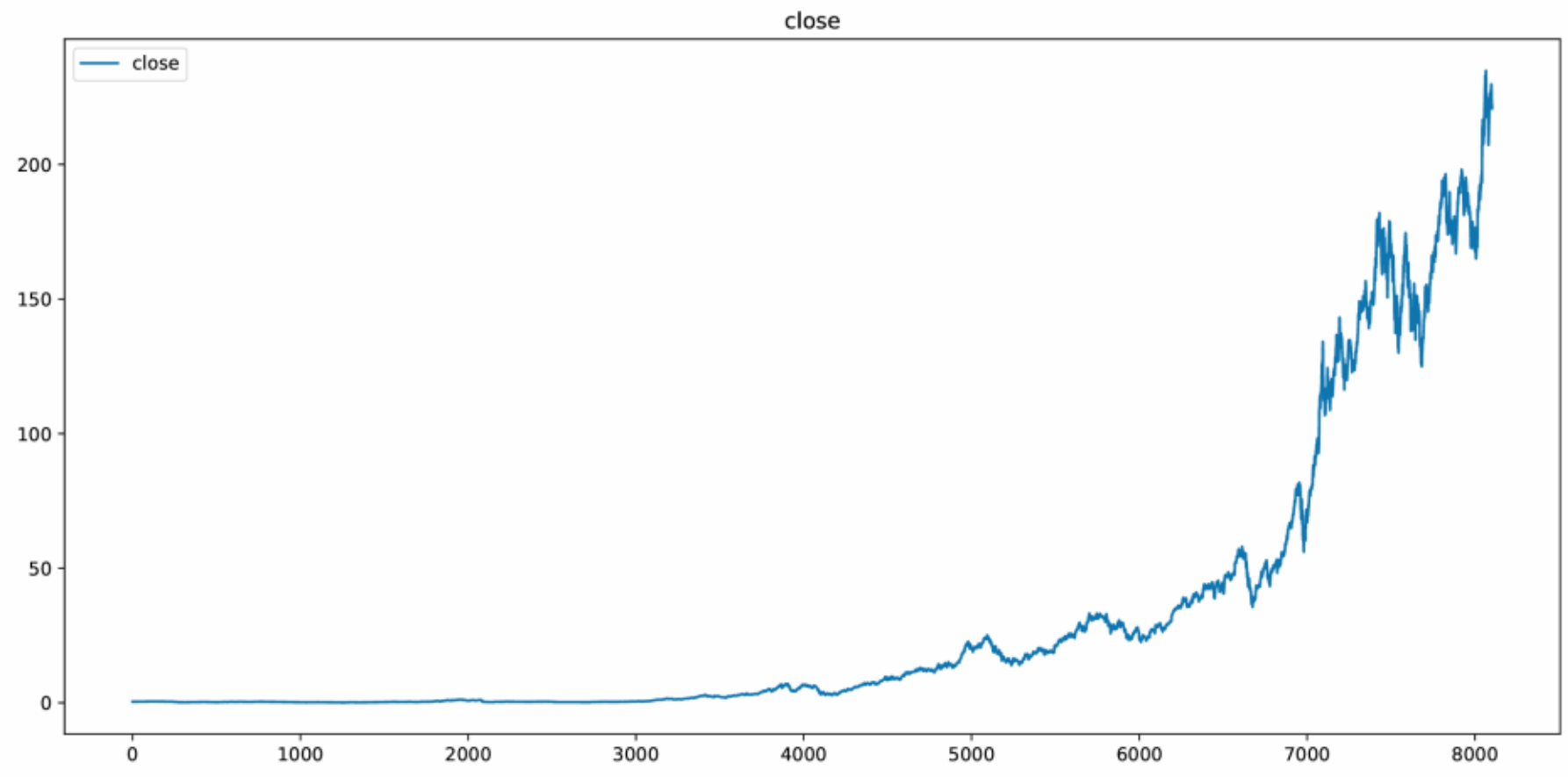

*Figure 6: Example of heteroskedastic, trending, non-stationary time series, the raw close price of AAPL*

**i. *Stationarity***

Non-stationarity can be either identified empirically or with the use of statistical tests like the Augmented Dickey–Fuller test (ADF) and the Kwiatkowski–Phillips–Schmidt–Shin test (KPSS) [11] [12]. The ADF test identifies the existence of a unit root, which indicates a stochastic trend, while the KPSS test addresses the presence of deterministic trends in the data.

Common approaches for addressing non-stationarity include computing logarithmic returns [11] [13] and simple returns [11] of the original time series to achieve stationarity. Differencing is another prominent transformation method in time series analysis [11], though it's insufficient for the AAPL stock price as it exhibits an exponential trend as made evident by its graph.

With the scope of discovering more effective ways of achieving stationarity for our specific time series, we test a novel approach which involves calculating the ratio of the non-stationary time series with the EMA of closeprev. The EMA adjusts for the exponential component that most of the features follow, simultaneously addressing both the changes in mean and variance. Additionally, we experiment with the EMA ratio after differencing first, for potentially better results.

**ii. *Outlier handling***

As a tree-based model, LightGBM is generally robust to outliers. However, we will be testing this hypothesis by using both features with and without outlier handling.

Financial data is excessively volatile, rendering the number of outliers, even in a stock like AAPL, unusually high. Moreover, outliers often signal economic crises or extreme investor confidence and rapid growth. In either case this information is essential for the model to be able to respond to such future scenarios.

To identify outliers, we use Tukey's Interquartile Range (IQR) method [12]. IQR is the difference between the first quartile (Q1) and third quartile (Q3). Outliers are defined as values that lie outside of [Q1- $n$ × IQR, Q3 + $n$ × IQR], where the value of $n$ varies, depending on the desired level of outlier handling. We use a value ranging from 1.5 to 3, but for the MACD Histogram in which extreme values are sparse but particularly skewed, we use a value of 20.

Instead of removing the outliers from the data or filling in with the mean, this study introduces a novel approach to retain crucial information that outliers provide, while reducing the skewing effect. More specifically, we take the difference of their value with the outlier threshold and take its square or cubic root, before adding it back to the threshold to get the normalized outlier value, we can repeat this process multiple times with different $n$ values. The mathematical equations for the calculation of the normalized outlier with square root can be seen below:

$$\text{new_outlier_value} = -\sqrt{|\text{outlier_value} - (Q1 - n \times \text{IQR})| + 1} + 1 + (Q1 - n \times \text{IQR})$$

*Equation 2: normalized outlier for below threshold values*

$$\text{new_outlier_value} = \sqrt{\text{outlier_value} - (Q3 + n \times \text{IQR}) + 1} - 1 + (Q3 + n \times \text{IQR})$$

*Equation 3: normalized outlier for above threshold values*

The reason we add 1 in the square root and then subtract it for above threshold values and add it back for below threshold values is because the square root function increases the input values which are less than 1, which would result in higher values for outliers the distance of which to the threshold is less than 1.

By reducing the distance of outliers with the rest of the data as well as with each other, we aim to facilitate LightGBM in the process of finding more meaningful splits for extreme values, without losing important information.

#### iii. *Standardization*

Data standardization is another data processing step which can be carried out after the stationarity transformations [14]. It is calculated by subtracting the mean and dividing with the standard deviation, ensuring that the new standardized time series has a mean of zero and standard deviation equal to one. By standardizing, all features acquire the same scale making it easier for LightGBM to identify patterns. In this study tests will be conducted both with and without standardization so as to evaluate its effectiveness.

## 6. Experiments

To evaluate the performance of LightGBM with the proposed features as well as compare each transformation method, we carry out multiple tests on the data set described in *Methodology.* We use an 80%-20% train-test split of the dataset.

### *6.1 Transformation Methods*

To begin with, we perform outlier handling to the duplicates of a specific set of features, so that we may compare their performance later on. Different transformation methods, described in *Methodology,* are iteratively applied to every feature whenever applicable, in order to create a new dataset, containing all the features in every transformation variant.

a. We apply simple returns to all features. Features with outlier handling only undergo this transformation to spare computational power.
b. We apply logarithmic returns to features that don't have negative values and features. The logarithm of negative values is not defined; therefore, we compute the cubic root instead.

c. The EMA ratio can only be applied to features that follow a similar structure to that of price in order to achieve stationarity.
d. The EMA difference ratio is applied to the same features as the EMA ratio

By combining all of those features a new data set is formed (DS1). We create another data set (DS2) as well by standardizing DS1. KPSS and ADF tests are carried out with significance levels of 0.05 on the entirety of DS1 and DS2 to ensure stationarity. In order to test the impact of different target variable transformation methods on model performance in relation to each different feature transformation method, we create 7 datasets from DS1 and DS2, each containing all features but different target variable transformations:

1. Simple Returns based on DS1
2. Logarithmic Returns based on DS1
3. EMA ratio based on DS1
4. EMA difference ratio based on DS1
5. Standardized Simple Returns based on DS2
6. Standardized Logarithmic Returns based on DS2
7. Standardized EMA ratio based on DS2

*Notes: Due to time constraints, the standardized EMA difference ratio was not tested. However, results from similar transformations provide a reasonable basis for comparison*
*The standardization process in this study was performed on the entire dataset (training and testing set combined) which results in a minor data leakage the impact of which is negligible. Additional experiments were conducted using standardization based solely on training set statistics. The results were consistent with those obtained using the full dataset statistics, confirming the previous assertion.*

By conducting this number of tests, we uncover relationships between different transformation methods as features and as target variables, aiming to identify the optimal combination for accurate and efficient stock market forecasting.

### *6.2 Training & Cross Validation*

With the scope of robustly generalizing the model, we opt for a rolling train-validation split, using TimeSeriesSplit from scikit-learn to avoid any data leakage. First, we split the training set into four segments to create three validation folds. In the first fold the model is trained on the first segment and validated on the next. The second fold uses the first two segments for training and the third for validation. Finally, the third fold makes use of all but the last segments for training, validating on the fourth.

### *6.3 Custom Loss Function*

As larger errors can result in detrimental consequences in the context of stock market forecasting [14], we opt for a custom loss function which penalizes them more aggressively.

### *6.4 Hyperparameter tuning*

Grid Search, commonly used in prior studies [15], is computationally expensive and time intensive. Instead, we employ Optuna's Bayesian Optimization algorithm, a faster and more scalable alternative. Balancing training efficiency with performance, we will be using 500 trials. Some of the parameters optimized can be seen below:

*Table 2: Hyperparameter optimization settings*

| *Parameter* | *Distribution type* | *Search Space* |
|---|---|---|
| *num_iterations* | integers | [500, 2200] |
| *learning_rate* | log uniform | [1e-5, 0.02] |
| *num_leaves* | integers | [10, 80] |
| *max_depth* | integers | [-1, 30] |
| *lambda_l1* | log uniform | [1e-8, 1e-3] |
| *lambda_l2* | log uniform | [1e-6, 10] |
| *max_bin* | integers | [125, 750] |

### *6.5 Metrics*

Accurately evaluating model performance is essential to derive meaningful results. For every target variable transformation method, we reverse transform the predictions accordingly in order to compare them to real market values. We calculate the Mean Absolute Error (MAE), the Root Mean Squared Error (RMSE), Directional Accuracy (DA) and training time. All testing was carried out on an AMD Ryzen 5 3600, with 32GB of DDR4 RAM, although it wasn't fully used by the training process. Furthermore, some computations were assigned to the Graphics Processing Unit (GPU) through CuPy to reduce load from the CPU.

### *6.6 Benchmarks*

The Random Walk (RW) model poses that all information about the next time point's value is reflected in the current value. It is the most fundamental benchmark for stock market forecasting as any model unable to outperform it essentially holds no meaningful predictive power. The equation for the RW model can be seen below:

$$P_t = P_{t-1} + \epsilon_t$$

*Equation 4: RW model*

- $P_t$: Price at time $t$
- $P_{t-1}$: Price at time $t-1$
- $\epsilon_t$: Random error term, assumed to follow $\epsilon_t \sim N(0, \sigma^2)$ (normally distributed with mean 0 and variance $\sigma^2$).

To improve workflow efficiency, we remove the random error term as it doesn't affect performance.

An ARIMA model was also considered but ultimately omitted as its accuracy improvement in relation to that of the RW model was negligible.

In addition to the RW model, we employ two LightGBM benchmark models. We remove the proposed features from DS1 and DS2 to create the third (DS3) and fourth (DS4) data set containing only conventional features, for comparison. The features removed can be seen below:

a. All features created with EMA ratio or EMA difference ratio
b. Slope Differences
c. ATR/open Ratios

Features with outlier handling as well as price gaps were not removed. LightGBM was trained on DS3 and DS4 the same way it was trained on DS1 and DS2 and the same metrics were calculated.

*Table 3: Tests summary*

<table>
<tr><th>Dataset</th><th>Features</th><th>Number of tests</th><th colspan="12">Target variable transformation methods used in tests</th></tr>
<tr><td>DS1</td><td>All</td><td>4</td><td colspan="3">Log Returns</td><td colspan="3">Returns</td><td colspan="3">EMA Ratio</td><td colspan="3">EMA Difference Ratio</td></tr>
<tr><td>DS2</td><td>All (standardized)</td><td>3</td><td colspan="4">Standardized Log Returns</td><td colspan="4">Standardized Returns</td><td colspan="4">Standardized EMA Ratio</td></tr>
<tr><td>DS3</td><td>Conventional</td><td>1</td><td colspan="12">Log Returns</td></tr>
<tr><td>DS4</td><td>Conventional (standardized)</td><td>1</td><td colspan="12">Standardized Log Returns</td></tr>
</table>

## 7. Results

The metrics extracted from the nine tests are denoted in the following table:

*Table 4: Metrics across tests*

| Target Variable Transformation Method | Training Time (hours, minutes) | DA | MAE | RMSE |
|---|---|---|---|---|
| Log Returns | 1h 38m | 63.15% | 1.3336 | 3.6475 |
| Standardized Log Returns | 6h 57m | 62.41% | 1.3422 | 3.6740 |
| Returns | 1h 24m | 63.64% | 1.3369 | 3.6859 |
| Standardized Returns | 7h 13m | 63.58% | 1.3435 | 3.7036 |
| EMA Ratio | 1h 49m | 58.02% | 1.3424 | 3.6277 |
| Standardized EMA Ratio | 3h 31m | 56.98% | 1.3507 | 3.6743 |
| EMA Difference Ratio | 1h 33m | 63.09% | 1.3365 | 3.665 |
| Benchmark (Log Returns) | 9h 9m | 60.43% | 1.4194 | 4.0677 |
| Benchmark (Standardized Log Returns) | 6h 49m | 60.99% | 1.4309 | 4.1489 |
| Random Walk | - | 50% | 1.6225 | 5.4932 |

We can infer a plethora of valuable findings from the results. To begin with, all models utilizing the proposed features achieve significantly higher accuracy both in terms of MAE and RMSE. As to DA, EMA ratio and its standardized counterpart significantly underperform, even losing to both the benchmarks. We speculate this is occurring because returns show how price is directly related to its previous, while EMA ratios show the change of price in relation to 14 other price points, which provides more insight about the trend but smooths out information about the direction of price change between two time points. This is supported by the fact that EMA Difference Ratio does not exhibit diminished DA. Differencing first solves the issue by forcing the model to predict the change in price for every time point, contributing positively to DA. We also notice that Returns show a minor increase in RMSE compared to Log Returns. This can be attributed to the presence of more extreme values in the target variable of Returns, due to the absence of the logarithmic component, which proves more difficult to predict by the model. To better illustrate the residual patterns, we plotted the target variable graph of Returns, alongside residuals:

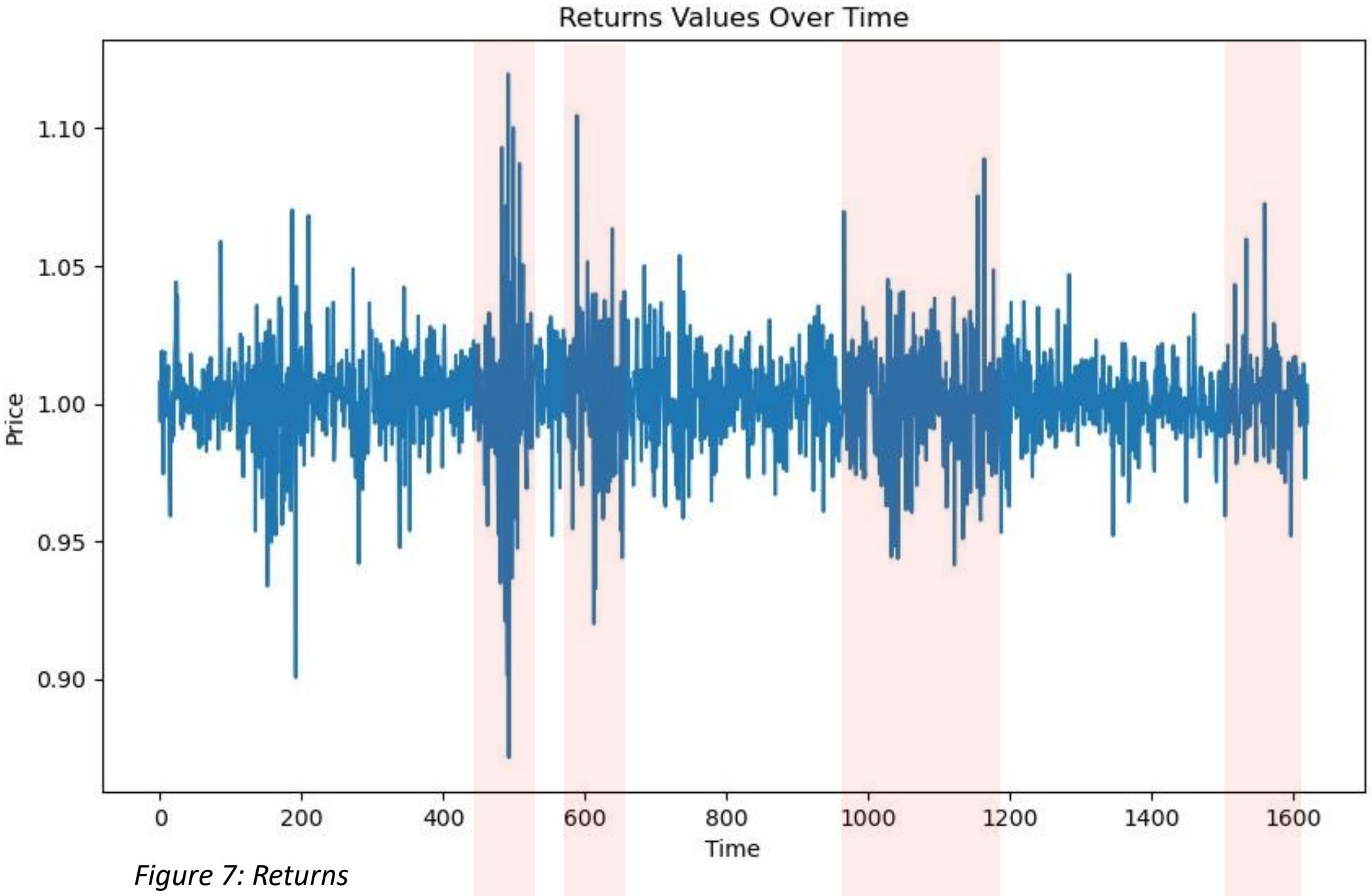

*Figure 7: Returns*

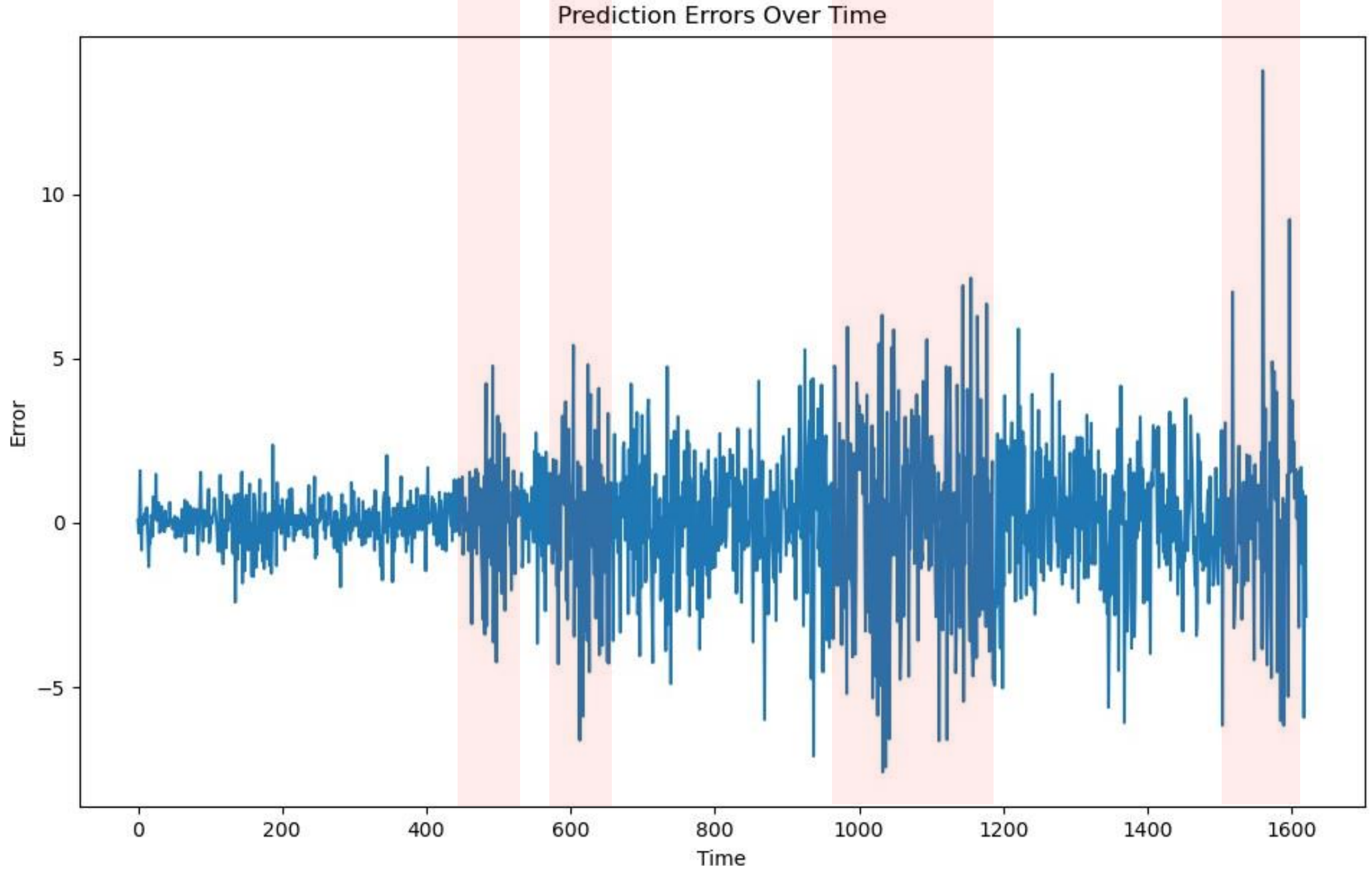

*Figure 8: Residuals for the model with Returns target variable transformation*

There are exceptions to the rule but generally clusters of high errors are accompanied by spikes in Returns values. Moreover, the graph indicates that performance can be further enhanced, as in an ideal scenario the residuals are independent and identically distributed (IID), whereas in the model, as made evident in the graph, the residuals are higher during certain periods.

Another interesting result is the training time which appears to be considerably higher in standardized transformations. More remarkably, standardization reduces training time in the benchmark. We hypothesize that standardization might increase numerical precision required by LightGBM to find optimal splits which could result in more complex computations. In addition, standardizing the novel features, like EMA ratios might distort the underlying scale, leading to less direct interpretability for the model's splits which would in turn require more computational time. In contrast, standardizing more conventional features may actually improve interpretability for the model, leading to decreased training time.

In terms of how the transformation methods compare, Log Returns followed by Returns and then EMA Difference Ratio achieve the best performance overall, although not deviating notably from the other transformation methods. However, they do outperform the random walk and the benchmarks by a large margin as can be observed in the table below:

*Table 5: Relative performance metrics*

| *Target Variable Transformation Method* | *Relative performance increase against RW (%)* | | | *Relative performance increase from RW, against Benchmark (Log Returns), (%)* | | | *Relative Training Efficiency* |
|---|---|---|---|---|---|---|---|
| | DA | MAE | RMSE | DA | MAE | RMSE | $MAE^2$×Time(s) |
| **Log Returns** | 26.3 | 17.81 | 33.6 | 26.07 | 42.25 | 29.47 | 10,507 |
| **Returns** | 27.28 | 17.6 | 32.9 | 30.77 | 40.57 | 26.78 | 9,153 |
| **EMA Ratio** | 16.04 | 17.26 | 33.96 | -23.11 | 37.85 | 30.86 | 11,877 |
| **EMA Difference Ratio** | 26.18 | 17.63 | 33.28 | 25.5 | 40.81 | 25.95 | 10,045 |
| **Benchmark (Log Returns)** | 20.86 | 12.52 | 25.95 | 0 | 0 | 0 | 66,362 |

All models outperform the RW by more than 10% in every metric which implies improved forecasting ability. In fact, in all models but the benchmark, RMSE decreased by more than 30% highlighting the models' ability to mitigate the effects of large, unexpected movements such as crisis. Relative accuracy increase from the MAE of the RW, against the benchmark reaches 42.25% with Log Returns, meaning that the difference from the RW MAE increased by 42.25% moving from the benchmark to the Log Return transformation method with novel features. We calculate training efficiency using the product of squared MAE with training time in seconds in order to place more weight on accuracy. The benchmark is significantly less efficient than all the other models. We hypothesize that this is occurring because the model is struggling to identify meaningful splits as easily in the conventional feature set.

Feature importance provides key insight to how each novel feature contributed to model performance as well as how each transformation method performed as a feature. In Log Returns, Returns, EMA Difference Ratio and their standardized counterparts, difference_open-closeprev_ema which is the feature modeling overnight price gaps, stationarized with the EMA, is extremely important, with EMA Difference Ratios and Returns features following up. Slope differences, ratios and other cross features rank high in importance, consistently beating conventional features, which manifests their effectiveness. open_ema is by far the most important feature in EMA ratio and its standardized counterpart, with multiple EMA ratio features and Returns features following up. In the benchmark, we observe a notably more balanced feature importance distribution, with Return-based features achieving the highest importance. Outlier handled features did not perform significantly better than their respective counterparts, highlighting LightGBM's robustness to outliers. Lastly, Log Return features are consistently the least informative, often falling to zero feature importance which implies the logarithm may smooth out information essential for the model.

*Figure 9: Top 10 Feature Importance scores for Returns*

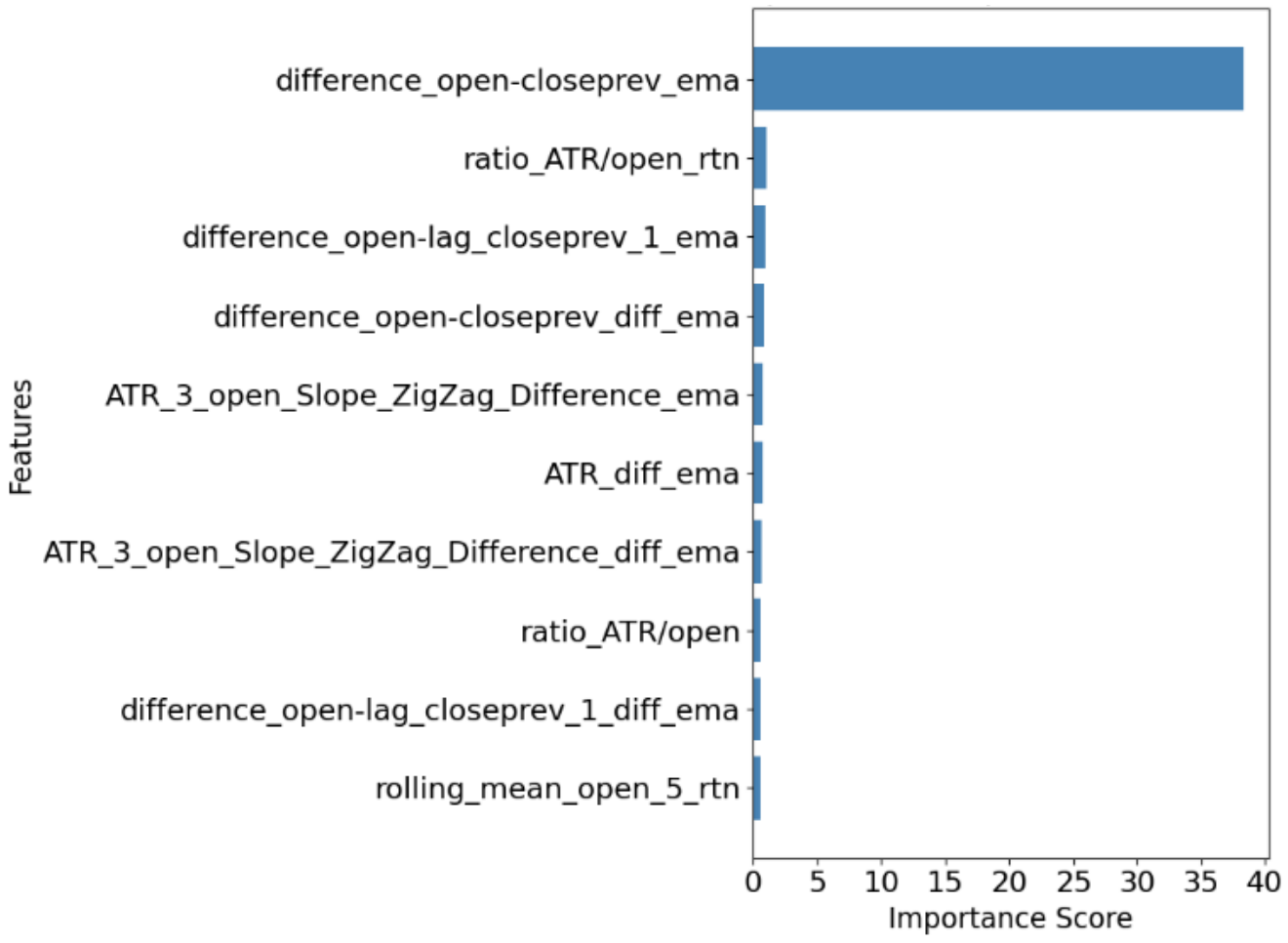

*Figure 10: Top 10 Feature Importance scores for EMA Ratio*

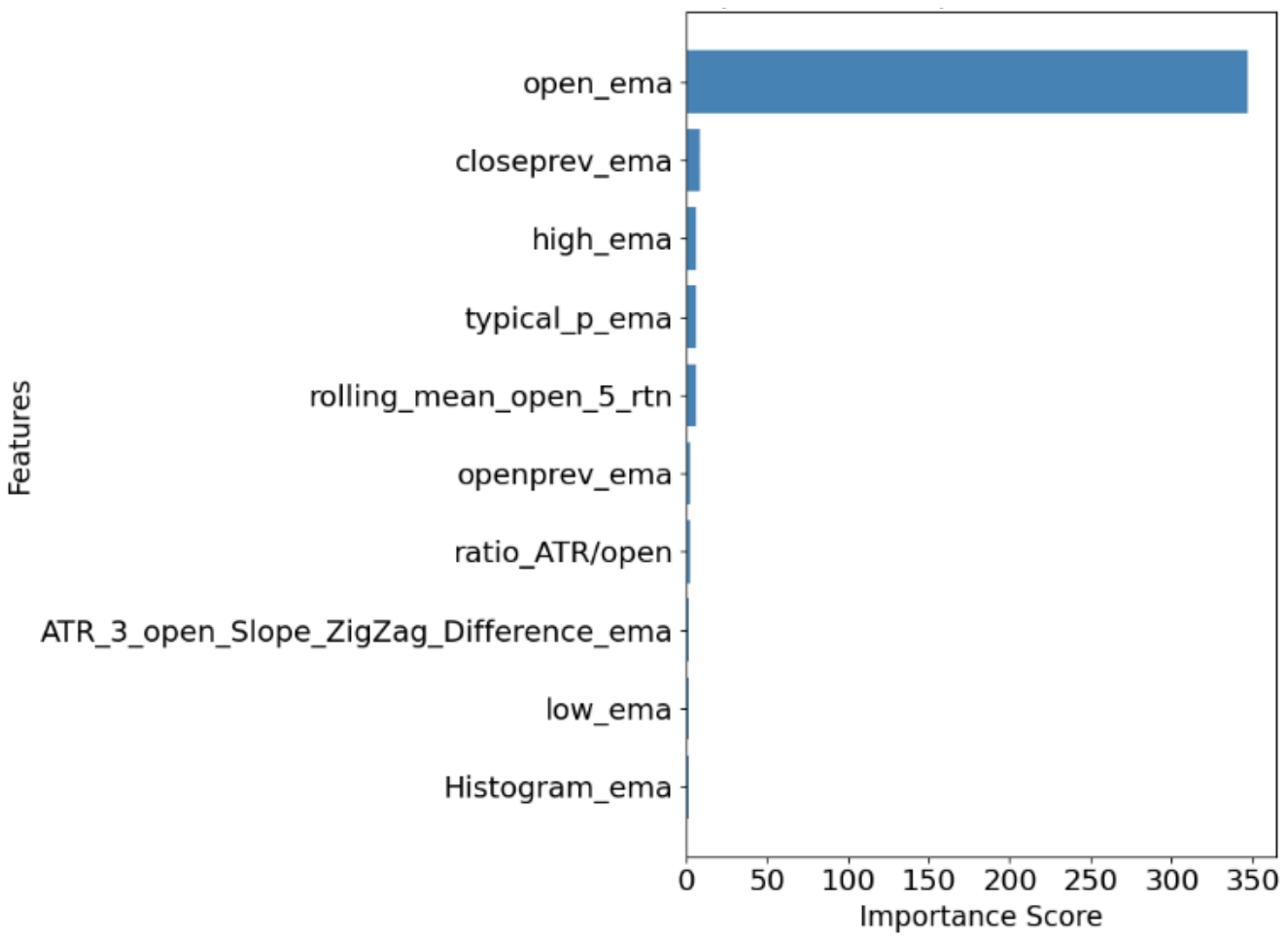

*Figure 11: Top 10 Feature Importance scores for Benchmark (Log Returns)*

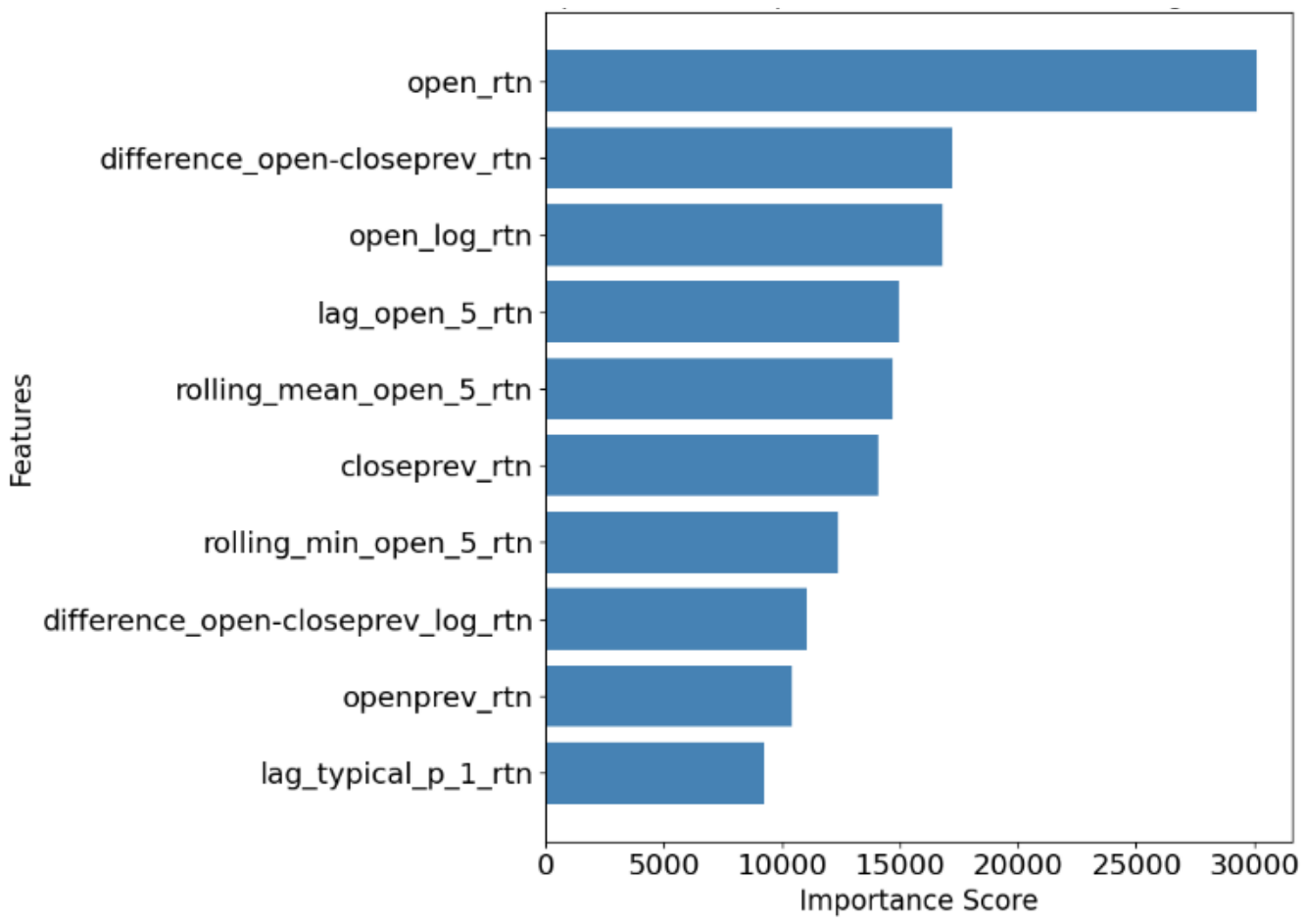

## 8. Conclusion

This study explored the use of novel feature engineering techniques in combination with innovative transformation methods to optimize the performance of LightGBM for stock market forecasting, testing on the AAPL daily prices. Different transformation methods were evaluated both on the target variable and on the features whenever applicable. For comparison, two conventional features-based LightGBM models and a RW model were deployed.

Our findings indicate that Log Returns were the optimal transformation method for the target variable, though the performance difference was generally minor. The standardized variants of transformation methods slightly decreased accuracy but greatly increased training time, underscoring that standardization is counterproductive for tree-based models like LightGBM.

The transformation methods drastically affected the importance of features on different models. Features transformed with EMA Ratios, EMA Differenced Ratios and Returns scored particularly high in feature importance throughout the models. Of crucial importance proved the novel features, especially the overnight price gap, consistently outperforming conventional features in terms of importance. This demonstrates their ability to provide the model with additional information about complex market dynamics, augmenting forecasting performance while exceedingly lessening training time.

The proposed outlier handling method doesn't offer much benefit for LightGBM, although testing was limited in that regard. Future experiments could better evaluate this method and its effectiveness in the broader context of time series forecasting. Additionally, residual plots indicated worsened model performance during high-volatility periods. Future work could involve training multiple specialized LightGBM models, each trained on a specific volatility regime. These regimes could be identified using a GARCH model or a LightGBM classifier. Lastly, EMA Ratio holds premise in multi-step ahead forecasting given its trend-based nature.

In conclusion, this study enhances the performance of LightGBM while reducing training time, through the use of novel features and transformation methods. Building upon these findings, future work could further improve accuracy, adaptability, and practical applicability of LightGBM and Machine Learning in financial time series forecasting.